\documentstyle[epsf]{mn-zi} 

\makeatother 
\def\figsize{\epsfxsize=\hsize}
\def\PD#1:#2/#3 {\if!#1!{\partial #2 \over \partial#3} 
      \else  {\partial^{#1} #2 \over \partial#3^{#1}} \fi} 
\def\eq#1{\begin{equation} #1 \end{equation}}

\let\mic=\micron 
\def\<#1>{\hbox{$\langle#1\rangle$}} 
\def\E#1{\hbox{$10^{#1}$}} 
\def\kms  {{\hbox{km s$^{-1}$}}} 
 
\def\about{{\hbox{$\sim$}}} 
 
\def\L  {{\hbox{$L_\ast$}}} 
\def\Lo {{\hbox{$L_\odot$}}} 
\def\M  {{\hbox{$M_\ast$}}} 
\def\Mo {{\hbox{$M_\odot$}}}

\def\arp {{\hbox{$a_{\rm rp}$}}}   
\def\ag {{\hbox{$a_{\rm g}$}}}  
\def\Tevap {{\hbox{$T_{\rm evap}$}}} 
\def\micron {{\hbox{$\mu{\rm m}$}}} 
\def\mic {{\hbox{$\mu{\rm m}$}}}  
 
\def\bullet{}

\begin{document}

\title{}
\author[]{}
\twocolumn    
\leftline{{\it To appear in:}}
\leftline{Proceedings of the The Sixth Annual } 
\leftline{Conference of the CFD Society of Canada,}
\leftline{June 7-9, 1998, Qu\'ebec, Canada}
\vskip 0.5in

\centerline{\Large {\bf  STELLAR OUTFLOWS DRIVEN  }}
\centerline{\Large {\bf  BY RADIATION PRESSURE }}
\vskip 0.2in

\centerline{\bf  \v{Z}. IVEZI\'{C}$^{\,1}$, G. R. KNAPP$^{\,1}$, M. ELITZUR$^{\,2}$ }
\vskip 0.1in

\centerline{\it $^1$Department of Astrophysical Sciences,}
\centerline{\it Princeton University, Princeton, NJ 08544}
\centerline{\it ivezic, gk@astro.princeton.edu}
\vskip 0.05in
\centerline{\it $^2$Department of Physics and Astronomy,}
\centerline{\it  University of Kentucky, Lexington, KY 40506}
\centerline{\it  moshe@pa.uky.edu}

\section*{ABSTRACT} 
We present a detailed, self-consistent model of radiatively driven  
stellar outflows which couples the radiative transfer and hydrodynamics 
equations. The circumstellar envelope, which consists of gas and dust, is 
described  as a two-component fluid to account for relative drifts. 
Our results agree with both molecular line observations and 
infrared continuum spectra, and show that steady-state outflows driven by 
radiation pressure on dust grains adequately describe the surroundings 
of the majority of cool luminous evolved stars.


\vskip 0.3in
\centerline{\large {\bf 1. INTRODUCTION }}   
\vskip 0.1in
Astronomical objects are often characterized by large linear scales, 
huge masses and powerful flows of energy.  Such environments emphasize 
different physical effects than those encountered in a laboratory or  
in engineering applications. An example is the radiation pressure 
force. This force results from the transfer of momentum from the  
radiation field to the intervening medium. For isotropic interaction  
between the radiation and the matter, the acceleration due to 
radiation pressure is ([1])  
\eq{ 
{{\bf a}_{rp}} = {1 \over c} \int^{\infty}_0 \chi_\lambda {\bf F}_\lambda 
   d \lambda 
}  
where $c$ is the speed of light, ${\bf F}_\lambda$ is radiative flux  
vector, $\chi_\lambda$ is the opacity per unit mass of the medium (including 
both absorption and scattering) and $\lambda$ is wavelength. In the  
engineering applications this force is usually negligible, but in some 
astronomical environments it can dominate over all other forces. 
 
Cool luminous evolved stars are one astronomical environment where radiation  
pressure plays a major role. These stars, which are in the last stages 
of their evolution, experience mass-loss. Although the exact nature  
of the mechanism that governs the initial ejection of mass from   
the stellar surface is still not clear [2], subsequent outflow is  
driven by the radiation pressure force [3,4]. When the expanding gas 
reaches a certain distance from the star it becomes sufficiently cool  
that dust grains (solid particles with sizes $\la 1 \micron$) begin to 
condense. Radiation pressure on the dust continues to push the material 
away from the star and a circumstellar shell, or envelope, is formed.  
Since the dust grains absorb stellar radiation and reradiate it in the 
infrared, the overall radiative spectral distribution is shifted toward 
longer wavelengths and many of these objects are invisible in the  
optical part of the spectrum. 
 
In order to solve the radiation transfer in such envelopes, the mass 
density distribution needs to be known. However, a given density  
distribution implies a particular velocity field which, in turn, is  
determined by the radiation pressure force. Thus, the radiative  
transfer and the dynamics are inherently coupled and  
have to be solved simultaneously. 
 
Our model assumes spherical symmetry and couples the momentum 
conservation and radiative transfer equations. The circumstellar 
envelope, which consists of gas and dust, is treated as a two-component  
fluid. We present an investigation of the dynamics and spectra of dusty 
envelopes around late-type stars, and discuss the implications for 
observations. 
 
The rest of this paper is organized as follows: first we discuss the
conditions for the radiation pressure force to be important (\S 2), then 
we describe our model (\S 3), and present the results (\S 4).  Some 
limitations of the model are discussed in \S 5.

\vskip 0.3in
\centerline{\large {\bf 2. WHEN IS THE RADIATION  }}   
\centerline{\large {\bf PRESSURE FORCE IMPORTANT?  }}   
\vskip 0.1in                         
  
Eq.$\,$ (1) shows that the acceleration due to the radiation pressure, 
\arp, is proportional to the radiation flux. This flux cannot 
be arbitrarily large because dust grains are radiatively heated,
and in sufficiently intense radiation field can evaporate. 
Thus the upper bound on the radiation pressure force is set not by
the intensity of the radiation field but by the evaporation 
temperature of the grains, $T_{\rm evap}$ \about 1000 K.

The radiation pressure force is maximized when the grains are heated by 
a directional radiation field, $|{\bf F}_\lambda|$ = c $u_\lambda$, where  
$u_\lambda$ is the radiation energy density. In this case, the total 
(bolometric) flux absorbed by grains when they are heated to \Tevap\ is: 
\eq{ 
    F \, \about\, \left({\Tevap \over 1000 \, {\rm K}}\right)^4 
      \, 10^5 \, {\rm W\, m^{-2}}. 
} 
In deriving the numerical constant we have assumed small grains with 
opacity proportional to $\lambda^{-1}$. With this flux, the acceleration 
due to the radiation pressure is 
\eq{ 
  a_{rp}^{max} \, \about \,  \left({\Tevap \over 1000 \, {\rm K}}\right)^4 
       \, {\rm m\, s^{-2}} \,\, \about \,\, 0.1 \, {\rm g},
} 
where g = 9.81 m s$^{-2}$. This is the {\it maximal} acceleration, 
that a grain can experience due to the radiation pressure force,
and is independent of the radiation field intensity.  
 
Dust grains are usually mixed with gas particles. In such a mixture 
the momentum which grains gain from the radiation field is  
transferred to gas particles via collisions [5]. In this case 
the effective acceleration of such a two-fluid system has to be  
multiplied by the dust-to-gas mass ratio ($\ll 1$), and thus 
\arp $\ll$ g. As a result, in engineering applications the radiation 
pressure force is always negligible compared to the gravitation force.   
 
When is the radiation pressure force important in astrophysical 
systems? The radiation pressure force will drive an outflow when it 
overcomes the gravitation force. Taking a typical circumstellar value 
for the dust-to-gas mass ratio of 0.002 [6], \arp \about 10$^{-4}$ g. 
Furthermore, this value can be even 10-100 times smaller when the dust 
drift effects become significant (this correction is proportional to 
$\rho_{\rm gas}^{1/2}$, and puts a lower limit on the mass-loss rate [2],
see \S3.1). 

The gravitational acceleration is 
\eq{ 
     \ag = G {M \over r^2}, 
} 
where $G$ is the gravitational constant, $M$ is the stellar mass, 
and $r$ is the distance from the star. Dust grains are heated  
to \Tevap\ at distance 
\eq{ 
  r \, \about \, 3 \times 10^{10}\,{\rm m}\,  \left({\L \over \Lo}\right)^{1/2} 
            \left({1000 \, {\rm K} \over \Tevap}\right)^2, 
} 
where $\L$ is the stellar luminosity, and \Lo = $3.86 \times 10^{26}$ 
W m$^{-2}$ is the  solar luminosity. The gravitational acceleration at 
the dust evaporation radius is thus 
\eq{ 
      \ag \, \about \, 0.01 \, g \, {M \over L} \,
            \left({\Tevap \over 1000 \, {\rm K}}\right)^4, 
}                                          
where $M$ = \M/\Mo\ (\Mo = 2 $\times$ 10$^{30}$ kg is the solar mass),  
and $L$ = \L/\Lo. Note that both \arp\ and \ag\ scale with $T_{\rm evap}^4$.

As the stellar luminosity increases the dust evaporation radius is
pushed further out, and \ag\ decreases.  For sufficiently small
$M$/$L$, \ag\ becomes smaller than \arp, and the radiation pressure
force drives an outflow of dust and gas. Taking a typical value of $M
\about 1$, the minimum required luminosity is 10$^3$ to 10$^4$ \Lo.
This is why a significant luminosity increase during the late stages of
stellar evolution is accompanied by outflows; it is the small local
\ag, rather than the large \arp, which enables the radiatively driven
stellar outflows.

\vskip 0.3in
\centerline{\large {\bf 3. A MODEL FOR RADIATION  }}   
\centerline{\large {\bf  DRIVEN OUTFLOW }}   
\vskip 0.1in

{\leftline{\bf 3.1 The Mathematical Model}} 

In this section we briefly present the basic equations describing the
mathematical model. Detailed discussion of the model and the
simplifications can be found in [2,7].

The equation of motion for the gas is \eq{\label{vg} v {dv \over dr} =
{1 \over c} \int^{\infty}_0 \chi_\lambda F_\lambda
  d \lambda  - { G M_{\ast} \over r^2} } where $v$ is the gas velocity.
The first term on the right-hand side of eq.$\,$(\ref{vg}) describes
the radiation pressure force, the second term is the gravitational
force exerted by the star. The force due to the gas pressure gradient
is omitted since it is negligible for the highly supersonic outflows
described here. In cool stars, the radiation pressure force acts only
on the dust grains (gas opacity is negligible).  Since the dust flows
out faster than the gas, the dust opacity decreases in proportion to
the ratio of the gas and dust outflow velocities \eq{
     \chi_\lambda \propto {v \over v + v_{drift}}, } where the dust
drift velocity with respect to the gas flow is determined from \eq{
     v_{drift} = \sqrt{Q_F \L v \over \dot{M} c}.  } Here $Q_F$ is the
flux averaged absorption efficiency, and $\dot{M}$ is the mass-loss
rate (for a detailed discussion of dust drift see [7]).

At first it appears from eq.$\,$(\ref{vg}) that outflows with arbitrary
mass-loss rates can be driven by the radiation pressure force.
However, the allowed mass-loss rate is limited from both below and
above. The lower limit is set by the above mentioned drift effects:  as
the mass-loss rate, i.e. gas density, decreases, the dust-gas coupling
decreases and with it the effective opacity. In the limit of large
mass-loss rates, practically the entire stellar radiation is absorbed
by dust, and shifted to longer, infrared wavelengths. As opacity is in
general a decreasing function of wavelength, the effective opacity
decreases with mass-loss rate in this regime, resulting in an upper
limit on the mass-loss rate.

\vskip 0.1in                                                              
{\leftline{\bf 3.2 The Solution Method }}   

The coupled system of the dynamics and radiative transfer equations 
is solved by an iterative scheme. For a given velocity field, the  
radiative transfer is solved separately by an exact scheme described 
in [8]. With known radiation flux, an updated velocity field is  
calculated from eq.$\,$(\ref{vg}), and these steps are repeated until 
both the radiation and velocity fields converge. As an initial
velocity field we take an analytic approximation obtained by
neglecting drift and reddening effects, and gravitational force [9].
 
The radiative characteristics of the star are largely irrelevant for  
the solution. We assume that the star radiates as a black body with  
$T_\ast = 2500$ K, typical for late-type red giant stars. Similarly, 
we assume that the dust evaporation temperature \Tevap = 1000 K. The 
dust optical properties are important because they determine the detailed
signature of the emitted spectrum. In this work we consider the two major 
grain types  that have been proposed for late-type stars.  Dielectric 
coefficients for ``astronomical silicate'' are taken from Ossenkopf et 
al. [10], and for amorphous carbon from Hanner [11]. A more detailed 
study including 5 different grain types and various mixtures is presented
in [4]. Absorption and scattering efficiencies are evaluated using 
Mie theory, assuming the dust grains are homogeneous spheres following
the size distribution proposed by Mathis et al. [12]. 
Carbon grains have smooth opacity, decreasing with wavelength, while
astronomical silicate has characteristic peaks at 9.7 \mic\ and 18 \mic.

\vskip 0.3in
\centerline{\large {\bf 4. RESULTS       }}   
\vskip 0.1in     
 
\leftline{{\bf 4.1  Scaling Properties of the Solution}} 
\begin{figure}
\centering \leavevmode \figsize \epsfbox[70 100 525 460]{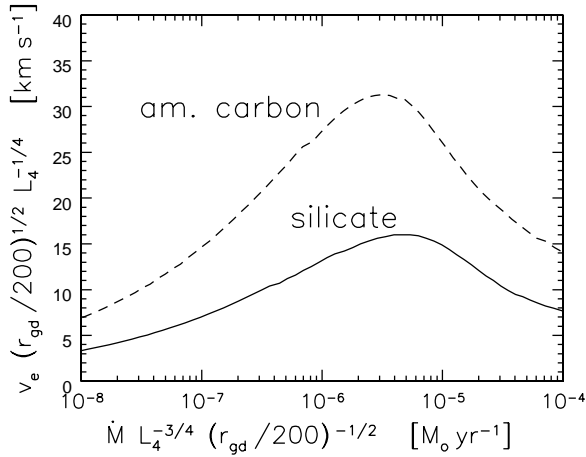}
\caption {Model results for the dependence of outflow velocity on mass-loss
rate $\dot{M}$, luminosity \L=$L_4 10^4$ \Lo, and gas-to-dust ratio r$_{\rm gd}$.
The plotted variables, whose forms follow from the model scaling properties 
and dimensional analysis, describe families of self-similar solutions for
two types of dust grains, as marked.} 
\end{figure}
For given dust optical properties, stellar temperature, dust
evaporation temperature, gas-to-dust ratio, mass-loss rate, and
luminosity, our model determines the outflow velocity field.  Optical
properties and the two temperatures are usually well constrained, so
in the remainder of this paper we will consider gas-to-dust ratio 
r$_{\rm gd}$, $v_e$, $\dot{M}$, and \L\ as the relevant quantities 
describing a model.  

The velocity field can be described by terminal outflow velocity $v_e$,
which sets the scale, and shape $u$ = $v/v_e$. 
From the scaling properties of the system, $v_e$ can be expressed as 
the product of a velocity scale proportional to $\L^{1/4} r_{\rm gd}^{\,-1/2}$, 
and a dimensionless function of the dust optical depth [9].
Because the solution is characterized by a single parameter, it follows 
from dimensional analysis that the
terminal outflow velocity is related to the input quantities through
some (quite involved) function $\Phi$ such that
\eq{
  v_e = \L^{1/4} \, r_{gd}^{\,-1/2}\,\,\, \Phi({ \dot{M} \over  \L^{3/4}
  r_{gd}^{1/2}}) 
} 
The argument of function $\Phi$ follows from the dependence of optical
depth on $\dot{M}$, \L, and $r_{\rm gd}$. That is, for given grains, 
there is a single family of self-similar solutions, $\Phi$, which 
relates $v_e r_{gd}^{1/2} \L^{\,-1/4}$ to 
$\dot{M} \L^{\,-3/4} r_{gd}^{\,-1/2}$. Strictly speaking, $\Phi$ also
depends on \M, and $\epsilon$ = $v_1 / v_e$, where $v_1$ is the velocity 
at the inner boundary, r$_1$, but detailed model calculations show that this 
dependence is negligible for all practical purposes. Solutions for two types 
of grains considered in this work are shown in figure 1 (obtained for 
$\epsilon$ = 0.1 and \M = 1 \Mo). These solutions can be used to reconstruct 
the $v_e(\dot{M})$ relationship for arbitrary \L\ and $r_{gd}$.

\begin{figure}
\centering \leavevmode \figsize \epsfbox[60 50 515 700]{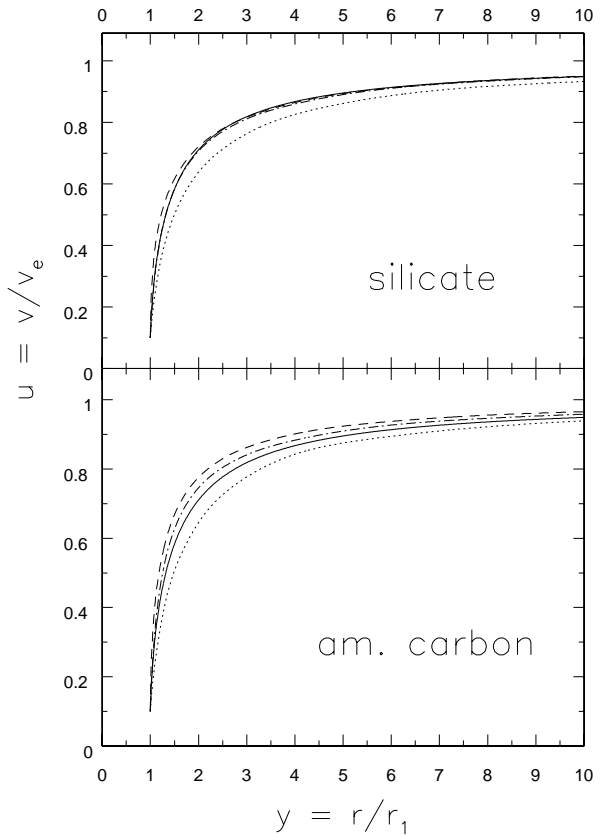}
\caption {Model results for the shape of velocity field $u = v(r)/v_e$
for two types of dust grains, as marked. Different curves correspond to 
different dust optical depths at 0.55 \mic\ (dotted lines:1, dot-dashed lines: 
10,  dashed lines: 100).  
The solid line corresponds to an approximate solution 
given by eq.$\,$ (11) with $k$ = 2. } 
\end{figure}

We find from detailed models that the shape $u$ is well described by 
\eq{
 u = \left(\epsilon^k + (1-\epsilon^k) (1 - {r_1 \over  r})
 \right)^{1/k}.  
}
The characteristic power index $k$ is a weak function of dust optical 
depth (fig. 2), ranging from 1.5 in optically thin envelopes to \about 2.5 
in optically thick ones ($k$=2 corresponds to an idealized case with 
acceleration proportional to r$^{-2}$). Again, the dependence on \M\ and 
$\epsilon$ is negligible.

\vskip 0.18in
\leftline{{\bf 4.2 Dynamical Properties of the Solution}} 
\begin{figure}
\centering \leavevmode \figsize \epsfbox[50 100 525 675]{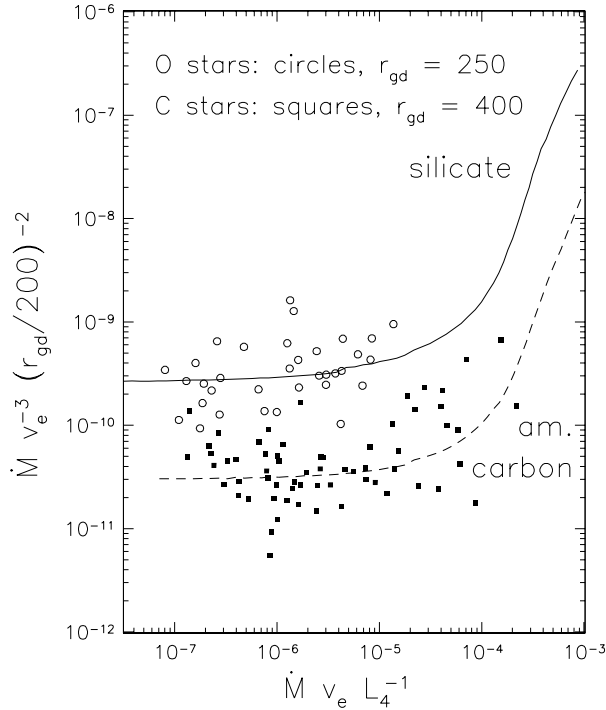}
\caption {Comparison of $v_e$ and $\dot{M}$ obtained from molecular line 
observations (symbols) with model predictions (lines). For more details
please see 4.2.} 
\end{figure}
The predicted relationship between the terminal outflow velocity $v_e$, and 
mass-loss rate $\dot{M}$, can be used to compare the model to observations.
Observations of molecular line emission (both thermal and maser) directly 
measure $v_e$, and with the aid of suitable emission models constrain $\dot{M}$.
Straightforward comparison of model predictions with observed $v_e$ and
$\dot{M}$ is hampered by uncertain values of \L\ and $r_{gd}$. Consequently, 
a direct utilization of the results presented in figure 1 is impractical
because effects of changing \L\ and $r_{gd}$ are coupled together. Their 
impact can be disentangled by comparing the observed and predicted 
values of $\dot{M} v_e^{\,-3} r_{gd}^{-2}$ and $\dot{M} v_e \L^{-1}$.
Such a diagram is shown in figure 3. Lines are model predictions and 
symbols are the observed values for two samples of stars: oxygen rich
(O) stars which are associated with silicate dust, and carbon rich
(C) stars associated with carbon dust. We have assumed that all stars
have $L_4 = \L / 10^4 \Lo = 1$, and constrained $r_{gd}$ by requiring
best agreement (in a least-squares sense for the whole sample) between the 
modeled and observed values of $\dot{M} v_e^{\,-3} r_{gd}^{-2}$. 
Best values are 250 for O stars, and 400 for C stars.
It is evident from figure 3 that these values are practically independent
of assumed value for luminosity because different values of \L\ shift
the data points only horizontally. 

The obtained values for $r_{gd}$ are in agreement with other independent
estimates (e.g. [6]). Close agreement between the model predictions
and observations provides strong evidence that outflows around 
late-type stars are radiatively driven. Additional evidence is provided
by the analysis of dust infrared emission.

\vskip 0.08in 
\leftline{{\bf 4.3 Infrared Emission}} 
\begin{figure}
\centering \leavevmode \figsize \epsfbox[30 72 535 705]{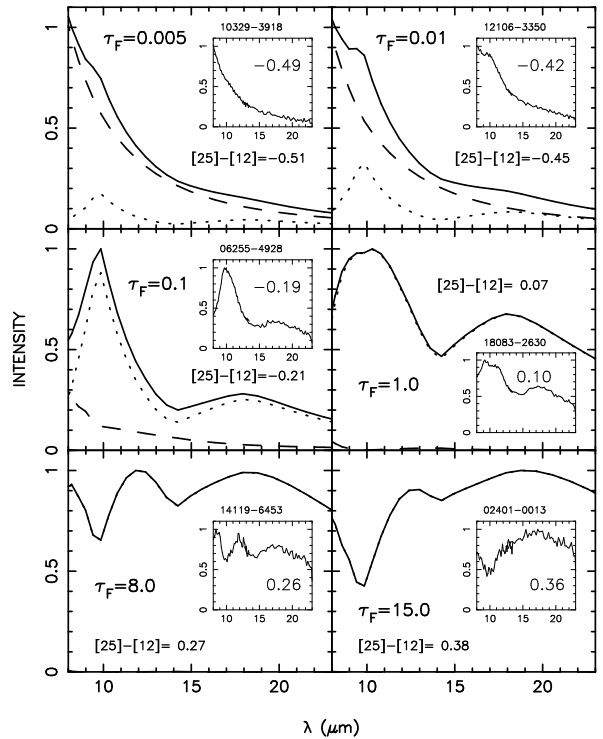}
\caption {Model spectra (normalized $\lambda F_\lambda$) for astronomical silicate 
grains for a sequence of flux averaged optical depths $\tau_F$.  Solid lines 
present the spectral shape of the emergent radiation, the sum of two components
--- the dust emission, indicated in dotted lines, and the emerging stellar radiation,
the dashed lines.  The model prediction for the IRAS [25]--[12] color is indicated. 
In each panel, the inset displays the measured spectrum of an actual source whose 
IRAS number is marked above the top left corner.  The measured [25]--[12] color of 
the source is listed in the inset in large type.} 
\end{figure}
An important property of the spherically symmetric radiative transfer 
problem is that, for given grains, the dimensionless spectral shape
$f_\lambda = (\lambda F_\lambda) / F$, where 
$F = \int_0^\infty F_\lambda d\lambda$ is the bolometric flux, 
depends only on the radial 
density profile and overall optical depth [8]. The radial density 
profile for an outflow is fully determined by the velocity profile 
through the mass conservation relation. Since the velocity profile 
depends only on optical depth, it follows that for each grain type
there is a family of spectra parametrized by optical depth. 
That is, irrespective of their individual values, any combination
of relevant parameters produces the same spectral shape as long as the 
corresponding optical depth does not change.  

A detailed comparison of observations with model spectra is given in 
[4], and here we only present results for silicate grains as an 
illustration. The six panels of figure 4 display a series of spectra 
parametrized by the flux averaged optical depth
\eq{
    \tau_F = \int_0^\infty \tau_\lambda {F_\lambda \over F} d \lambda.
}

The plotted wavelength range matches the spectral observations obtained by 
the Infrared Astronomical Satellite Low Resolution Spectrometer (IRAS LRS)
[13].  Dashed lines represent the direct stellar radiation emerging
from the system, dotted lines are the dust emission and full lines are
the total emission --- the model predictions of observed spectra.  In
optically thin envelopes ($\tau_F < 1$) the strength of the 9.7
\mic\ silicate emission feature increases with optical depth until $\tau_F$
reaches \about\ 1.  With further increase in $\tau_F$, this emission
turns into an absorption feature whose depth is proportional to
$\tau_F$. As an illustration, the inset in each panel displays the
observed spectrum of an actual source whose IRAS number is indicated
above the top left corner.  Each panel also indicates the [25]--[12]
color\footnote {Colors are convenient parameters used in astronomy to 
describe spectral shapes. Here we define [25]-[12] as 
$\log({F_{25} / F_{12}})$, where $F_{12}$ and $F_{25}$ are fluxes at 12 \mic\
and 25 \mic, respectively.} of the computed model while the number listed 
in large type in the inset is the IRAS color of the displayed source. 
The close agreement of spectral evolution between model predictions and
observations is evident.

\vskip 0.3in
\centerline{\large {\bf 5. DISCUSSION    }}   
\vskip 0.1in     
Since our models assume steady state, they can only describe the
time-averaged behavior of the outflows.  This is an adequate
description as long as time variability occurs only on time scales
shorter than the averaging time.  Calculations show that most of the
radiative acceleration takes place within $r$ \about\ \E{16} cm.  With
typical outflow velocities of \about\ 10 \kms, this distance is covered
in \about\ 300 years, the relevant time scale for averaging density.  
Although all late-type stars display pulsational or irregular variability, 
those variations are characterized by time-scales of a few years at most,
much shorter than the averaging time.  Thus the assumption of steady
state seems justified.  Additional time dependence can be introduced by
chemical evolution of the stellar atmosphere, possibly leading to
gradients in the grain chemical composition.  Though our models do not
include such an effect, they adequately describe the envelopes as long
as chemical gradients occur only at $r \ga \E{16}$ cm.  The reason is
that the grain composition in that region is largely irrelevant since
all grains behave similarly at low temperatures and long
wavelengths.  And since spatial gradients with length
scale $\la$ \E{16} cm imply temporal variations with time scales $\la$
100 years, the durations of phases not covered by our models do not
exceed \about\ 100 years\footnote{It has been recently suggested that 
mass-loss rate for some stars might change on similar time scales [14].}.  
If typical lifetimes in the mass-losing phase are \about\ \E4--\E5 years, 
only \about\ 0.1--1\% of all sources
cannot be described by our models. Evolutionary effects characterized
by time scales longer than \about\ 100 years can be incorporated into
our models as adiabatic changes of the relevant sources.  

An important ingredient of our models is the assumption of spherical
symmetry.  Small departures from sphericity, such as a slight
elongation, should not affect the results significantly.  However, our
models do not apply in the case of major deviations from spherical
symmetry such as disk geometry or strong clumpiness with characteristic
size scale less than \about\ \E{16} cm.  The success of our models in
describing the infrared spectra and mean mass-loss rates 
indicates that such deviations may not be important for the majority of
late-type stars.

{\em ACKNOWLEDGEMENTS:} Support by NSF grant AST96-18503 is gratefully
acknowledged.

\vskip 0.3in
\centerline{\large {\bf REFERENCES }}   
\vskip 0.1in    
\begin{itemize}

\item \hskip -0.26in $[1]$ G. B. Rybicki and A.P. Lightman, {\em Radiative Processes
in Astrophysics}, John Wiley \& Sons, New York, 1979.

\item \hskip -0.26in $[2]$ M. Elitzur, in {\em Instabilities in evolved super- and
hypergiants,} eds. de Jager and Nieuwenhuijzen (Amsterdam), p. 60.,
1991.

\item \hskip -0.26in $[3]$ E.E. Salpeter, {\em The Astrophysical Journal}, Vol. 193,
585, 1974.

\item \hskip -0.26in $[4]$ \v Z. Ivezi\' c and M. Elitzur, {\em The Astrophysical
Journal}, Vol. 445, 415, 1995.

\item \hskip -0.26in $[5]$ R.C. Gilman, {\em The Astrophysical Journal},  
Vol. 178, 423, 1972.
  
\item \hskip -0.26in $[6]$ G.R. Knapp, {\em The Astrophysical Journal},  
Vol. 293, 273, 1985. 

\item \hskip -0.26in $[7]$ N. Netzer and M. Elitzur, {\em The Astrophysical Journal},  
Vol. 410, 701, 1993.  

\item \hskip -0.26in $[8]$ \v Z. Ivezi\' c and M. Elitzur, {\em Monthly Notices of the Royal 
Astronomical Society}, Vol. 287, 799, 1997. 

\item \hskip -0.26in $[9]$ \v Z. Ivezi\' c and M. Elitzur, {\em in preparation}, 1998.  


\item \hskip -0.26in $[10]$ V. Ossenkopf, Th. Henning, J.S. Mathis, {\em Astronomy \& Astrophysics},  
Vol. 261, 567, 1992.

\item \hskip -0.26in $[11]$ M.S. Hanner, {\em NASA Conf. Pub.},  
Vol. 3004, 22, 1988. 

\item \hskip -0.26in $[12]$ J.S. Mathis, W. Rumpl, K.H. Nordsieck, {\em The Astrophysical Journal},  
Vol. 217, 425, 1977.  
 
\item \hskip -0.26in $[13]$ IRAS Catalogs and Atlases, Atlas of Low Resolution IRAS Spectra,
       1986, Joint IRAS Science Working Group, prepared by F.M.
       Olnon \& E. Raimond, {\em Astronomy \& Astrophysics Supplement}, 65, 607.

\item \hskip -0.26in $[14]$ \v Z. Ivezi\' c and G.R. Knapp, {\em in preparation}, 1998.  

\end{itemize} 
 
\end{document}